\documentclass[12pt]{article}

\usepackage{amsmath,wrapfig,amssymb,fullpage,graphicx}
\usepackage{booktabs}

\begin{document}

\begin{titlepage}

\begin{center}
\LARGE{Mixture Likelihood Ratio Scan Statistic for Disease Outbreak Detection} \\
\vspace{12pt}
\large{Michael D. Porter$^{1,2}$, Jarad B. Niemi$^{3}$, and Brian J. Reich$^1$   } \\
\end{center}
{\small
 \begin{flushleft}
  $^1$ Department of Statistics, North Carolina State University, Raleigh, NC, USA. \\
  $^2$ Statistical and Applied Mathematical Sciences Institute (SAMSI), Research Triangle Park, NC, USA. \\
  $^3$ Department of Statistical Science, Duke University, Durham, NC, USA.
 \end{flushleft}
}
 \footnotetext{Correspondence to Jarad Niemi, Department of Statistical Science, Duke University, Durham, NC, 27708-0251. (jbn9@stat.duke.edu)}

\vspace{1in}
\noindent Running Head: Mixture Likelihood Ratio Scan Statistic

\end{titlepage}

\newpage
\section*{\normalsize{BASIC CONCEPT OF ALGORITHMIC APPROACH}}
\label{s:model}
Our method assumes a Poisson distribution for daily counts from each of the different data sources. We modeled the daily mean as a baseline plus an outbreak component. The main contribution of our approach is to explicitly model the outbreak component. Guided by the distinct outbreak signatures in the training data, we assumed parametric forms for the outbreak profiles. This allows the development of a set of mixture likelihood ratio statistics, one for each possible outbreak starting date. The Mixture Likelihood Ratio Scan Statistic (MLRSS) is the result of scanning over the possible starting dates to find the most likely one.

This approach, based on standard sequential change point methodologies, enjoys several properties. By taking into account the outbreak profiles, we are leveraging more information about the total process (both before and after an outbreak) than if we only considered nonspecific deviations from the ``in-control'' process. Furthermore, although not an objective of the contest, our method facilitates the estimation and prediction of the outbreak start time, severity, and length. This could be useful in planning and evaluating mitigation strategies once an outbreak is detected. Our method can be extended to multiple data sources, space-time surveillance, or multiple syndromes.
Finally, the computation is quick enough to allow this method to be used when the data arrives much more frequently than once per day.

\subsection*{\small{Description of model}}

For each data source, we assume $o_t$, the daily counts on day $t$, are generated independently as
 \[ o_t \sim Pois(\lambda_t + \delta_t(t_o,\theta)) \]
where $\lambda_t$ is the baseline mean for time $t$ and $\delta_t(t_o,\theta)$ is the mean excess due to an outbreak. The baseline mean has the form $\lambda_t = \exp(X_t' \beta)$ where the vector $X_t$ contains terms for day of week and seasonal effects and the vector $\beta$ are parameters that are estimated from the training data. If there is no outbreak, $\delta_t(t_o,\theta)$ is equal to zero and we denote the density of $o_t$ as $f_0(o_t)$. If there is an outbreak, the profile, $\delta_t(t_o,\theta)$, is a function of the start time of the outbreak, $t_o$, and shape parameters, $\theta$. The density of $o_t$ under an outbreak is denoted $f_1(o_t;\delta_t(t_o,\theta))$. 

\subsection*{\small{Outline of surveillance methodology}}

We found evidence in the training data of distinct outbreak signatures for each data source during an outbreak. To incorporate this information into our surveillance statistics we adopted a likelihood ratio based approach.

Let $\Lambda_{t_o}^t(\theta)$ be the likelihood ratio (LR) for an outbreak at time $t$ which started at time $t_o$ versus no outbreak. Under our assumptions, this LR is

\begin{equation}
\Lambda_{t_o}^t(\theta) \triangleq \prod_{s=t_o}^t \frac{f_{1}(o_s;\delta_s(t_o,\theta))}{f_{0}(o_s)} = \prod_{s=t_o}^t e^{-\delta_s(t_o,\theta)} \left( 1 + \frac{\delta_s(t_o,\theta)}{\lambda_s} \right)^{o_s}
\label{LR}
\end{equation}

While we assume knowledge of the parametric form of the outbreaks profiles, we do not know the exact parameter values for a given outbreak. One approach for dealing with this uncertainty is to integrate the likelihood ratio with respect to some probability  density $h$ of $\theta$. This approach, termed \textit{Mixture Likelihood Ratio} (MLR) (\cite{pollak},\cite{pollak/sieg}) creates a new statistic defined in \eqref{mlr}. If $h$ is a discrete uniform distribution with mass at $\{\theta_1, \theta_2, \ldots, \theta_{n_\theta}\}$, the MLR simplifies to the equation on the right side of \eqref{mlr}.
\begin{equation}
S_{t_o,t} \triangleq \int_{\theta \in \Theta} \Lambda_{t_o}^t(\theta) \, h(\theta) \, d\theta = \frac{1}{n_\theta} \sum_{j=1}^{n_\theta} \Lambda_{t_o}^t(\theta_j)
\label{mlr}
\end{equation}

For each time $t$, $S_{t_o,t}$ is calculated for every outbreak starting time $t_o$ in a window $W_t$. The window is set to limit the amount of past data considered for the starting time of the outbreak. By scanning over the possible outbreak start times, we obtain the \emph{Mixture Likelihood Ratio Scan Statistic} (MLRSS)
\begin{equation}
 R_t \triangleq \max_{t_o \in W_t} S_{t_o,t}
\label{mlrss}
\end{equation}

The MLRSS provides evidence that an outbreak has begun sometime \emph{prior} to the current time $t$. Therefore, this statistic will continue to take a large value, even after the outbreak period, as long as $W_t$  still contains the outbreak start time. However, the Technical Contest evaluates an algorithm score that assesses evidence that an outbreak is occurring \emph{at a particular time}. To get an appropriate algorithm score, we took the least squares slope estimate of $\{R_s: s=t-S, \ldots, t \}$ over the last $S+1$ days. Thus our algorithm score, $a_t$ took the form of a weighted sum where $w_s$ are the weights given to estimate the slope
\begin{equation}
a_t = \sum_{s=t-S}^t w_s R_s
\label{alg.score}
\end{equation}
This algorithm score is large when $R_t$ is increasing dramatically and around zero when $R_t$ is essentially constant.

\section*{\normalsize{ADAPTATIONS FOR THE CONTEST}} \label{aftc}
The characteristics of each day were described by variables in the vector $X_t$, which was comprised of a weekday/weekend indicator, sine and cosine functions of time with 1, 2, 4, 8, and 16 periods per year, and the interactions between the weekday/weekend indicator and the sine/cosine functions with 1 and 2 periods per year.  The $\beta$ coefficients were estimated from the training baseline data and held fixed in the subsequent analysis of the testing data.

\begin{figure}[ht]
\begin{center}
\includegraphics[scale=.5]{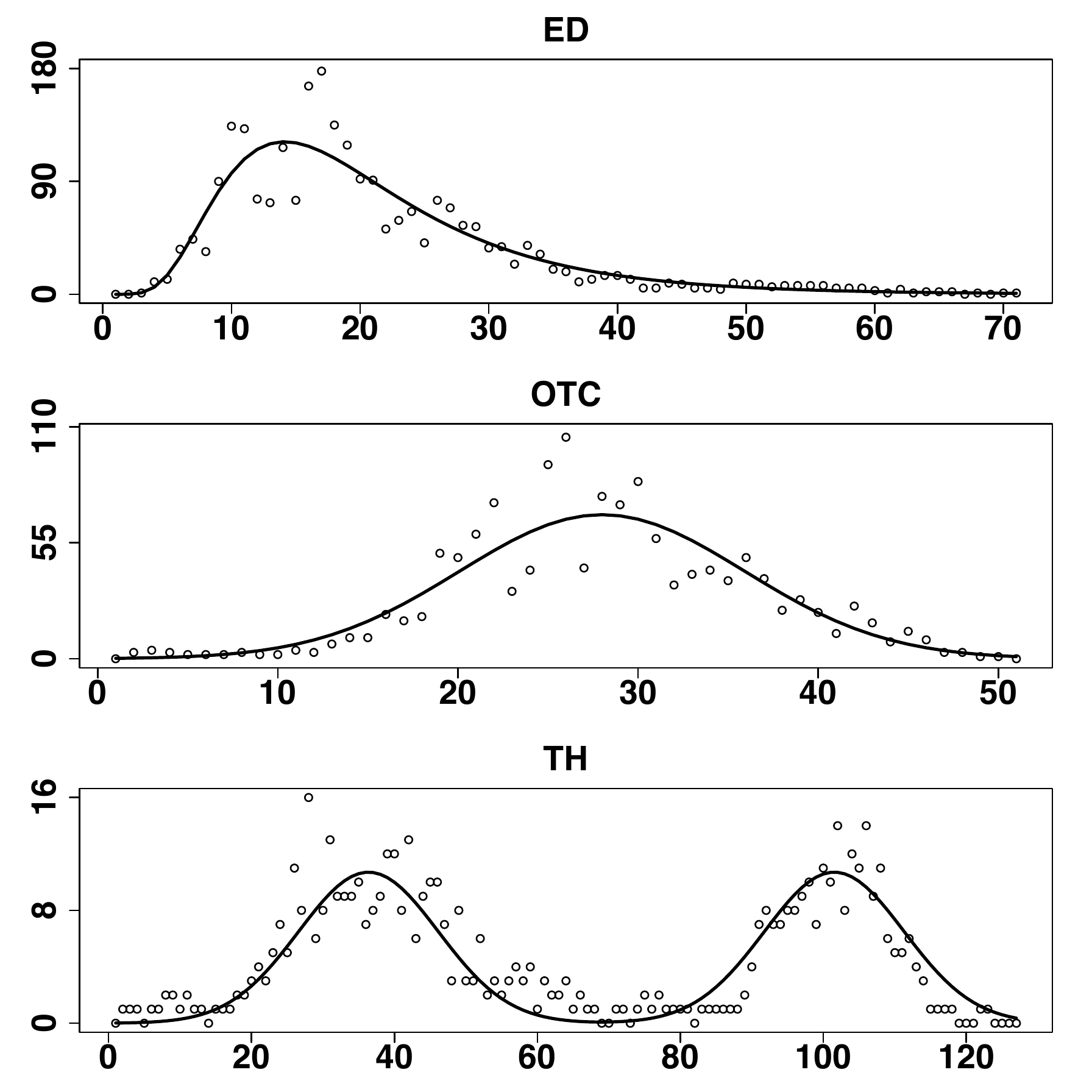}
\caption{Outbreak profiles (\emph{lines}) estimated for the first training outbreak (\emph{points}) of each data source. The $x$-axis is days into outbreak and the $y$-axis is the excess counts. }
\label{fig:outbreaks}
\end{center}
\end{figure}

The most important adaptation of our approach is determining the parametric form of the outbreak profiles, $\delta_t(t_0,\theta)$. For each data source, we extracted the 30 outbreak signatures by subtracting the common baseline count from the daily counts. Visual inspection of these outbreak signatures led us to assume three parametric outbreak profiles. Figure \ref{fig:outbreaks} shows the outbreak profile estimated for the first training outbreak of each data source.

The mathematical form of these outbreak profiles is shown in Table \ref{tab:outbreaks}. For the ED and OTC data sources
we used a log-normal and Gaussian kernel, respectively, with $\theta = (c,\mu,\sigma)$. The TH outbreak signatures where more complicated, often bimodal.  Therefore we used a two-component Gaussian mixture with $\theta = (c,\mu_1,\mu_2,\sigma)$. For all of these data types $c$ affects the severity of the outbreak, $\mu$ affects the peak day (or days) of the outbreak, and $\sigma$ the duration of the outbreak.

\begin{table}[b]
\begin{center}
\begin{tabular}{ll}
\toprule
Source & $\delta_t(t_o,\theta)$ \\
\midrule
ED & $c\exp(-(\log(t-t_o+1)-\mu)^2/\sigma)$ \\
OTC & $c\exp(-(t-t_o+1-\mu)^2/\sigma)$ \\
TH & $c\left[\exp\left(-\frac{1}{\sigma}[(t-t_o+1-\mu_1)^2+(t-t_o+1-\mu_2)^2]\right)\right]$  \\
\bottomrule
\end{tabular}
\caption{Mathematical form of the outbreak profiles.}\label{tab:outbreaks}
\end{center}
\end{table}

In addition to using the training data to determine the parametric form of the outbreak curves, we also used the 30 training outbreak signatures to calculate values for the outbreak parameters $\theta$. For the $j^{th}$  $(j=1,\ldots,30)$ training outbreak signature, we estimated
$\theta$ using maximum likelihood, giving $n_\theta=30$ potential curves
indexed by $\hat{\theta}_1,\ldots,\hat{\theta}_{30}$ for each data source. This provided a uniform discrete distribution over the possible values of $\theta$ which were used in \eqref{mlr} by plugging in $\theta_j=\hat{\theta}_j$.

Equation \eqref{mlrss} required a window $W_t$ to reduce unnecessary computation \cite{lai}. We used an adaptive window which depends on the most likely outbreak starting time estimated from the previous day. We used the window
\begin{equation*}
W_t = \{ \min (t^*,t-10), \ldots,  t-1 \}
\end{equation*}
where $t^*$ is the value of $t_o$ maximizing $S_{t_o,t-1}$. This window always includes at least 10 days, but will extend if the current estimate of the start time is further in the past.

In the algorithm scores for ED, OTC, and TH, we used $S=7$, $S=12$, and $S=10$ respectively in \eqref{alg.score}. The weights are given by
\begin{equation*}
w_s = \frac{x_s-\bar{x}}{\sum_{i=1}^{S+1} (x_i - \bar{x})^2}
\end{equation*}
where $x_i=i$ and $\bar{x} =  \left(\sum x_i \right)/(S+1)$.

\section*{\normalsize{IMPLEMENTATION DETAILS}}
For OTC and TH data, we found that this procedure outperformed several standard methods.  However, for ED data our approach gave similar results to the commonly-used and computationally-convenient exponentially weighted moving average (EWMA) approach.  Therefore, for ED data we used the  EWMA algorithm score $a_t = (1-\phi) a_{t-1} + \phi  r_t$, where $a_0 = 0$ and $r_t = \max(o_t - \lambda_t,0) / \sqrt{\lambda_t} $ is the truncated standardized residual. There were no attempts to address outliers for this data source and we used $\phi=0.25$.

For the OTC and TH data, we used an ad-hoc outlier remediation step in the calculation of the mixture likelihood ratios in \eqref{LR}. We looked for large deviations in the standardized residuals from a fitted outbreak profile. If
\[ \max_{s \in W_t} \frac{o_s - \left(\lambda_s+\delta_s(k,\hat{\theta}_j ) \right) }{\sqrt{\lambda_s+\delta_s(k,\hat{\theta}_j)}} > \gamma \]
for $\gamma=23$, we changed $o_{s^*}$ to $\lambda_{s^*}+\delta_{s^*}(k,\hat{\theta}_j)$, where $s^*$ is the time with the largest residual.

\section*{\normalsize{LESSONS LEARNED AND FUTURE RESEARCH DIRECTIONS}}

Methods that attempt to model the outbreak profile directly are sensitive to the information available concerning the profile's form. Therefore in real life situations these methods may be more appropriate for influenza and \emph{E. coli} outbreaks than they would be to anthrax. Epidemic modelers often build susceptible-infected-recovered (SIR) models for disease outbreaks that are based on infection and recovery rates. We are currently working on including these types of models into our method.

In real world application of this methodology assuming a known baseline mean will probably lead to inaccurate surveillance. Performing periodic checks for the accuracy of the baseline and incorporating uncertainty into the baseline parameters could improve overall detection performance.

\section*{\normalsize{CRITIQUE OF CONTEST METHODOLOGY AND SUGGESTIONS FOR FUTURE CONTESTS}}

Our first suggestion for future contests is to require the contestants to \emph{learn} the baseline in the testing data. Our experience suggests the parameters used to generate the training data baseline were the same or very similar to those used for the testing data baseline. A more realistic situation would require the contestants to simultaneously estimate the baseline and find outbreaks.

Our second suggestion is to modify the scoring to be a real-world cost function. The cost function should incorporate costs due to false alarms as well as costs due to the detection delay of a true outbreaks. In particular, there should be a difference in early versus late detection of the OTC and TH data. The scoring system for the ED data implicitly implies a cost function which is linear in detection delay. We suggest this relationship is not linear and that it may even depend on the outbreak severity.

Overall we were very pleased with this contest and enjoyed ourselves. Thank you for the opportunity to participate.

\section*{\normalsize{ACKNOWLEDGEMENTS}}
The research conducted by M. Porter and B. Reich has been supported in part by
National Science Foundation grant DMS-0354189.

\end{document}